\documentstyle[12pt,epsfig]{article}
\input axodraw.sty

\newcounter{minutes}
\def\lsim{\:\raisebox{-0.5ex}{$\stackrel{\textstyle<}{\sim}$}\:}

\newcommand{\newc}{\newcommand}
\newc{\pbi}{pb$^{-1}$}
\newc{\ie}{{\it i.e.} }
\newc{\ti}{\tilde}
\newc{\ra}{\rightarrow}
\newc{\ee}{$e^+e^-$\ }
\newc{\mm}{$\mu^+\mu^-$}
\newc{\taus}{$\tau^+\tau^-$}
\newc{\uu}{$u\bar{u}$\ }
\newc{\qq}{$q\bar{q}$\ }
\newc{\dd}{$d\bar{d}$\ }
\newc{\ud}{$u\bar{d}$\ }
\newc{\eeee}{$e^+e^-\ra e^+e^-$\ }
\newc{\eemm}{$e^+e^-\ra \mu^+\mu^-$\ }
\newc{\eett}{$e^+e^-\ra \tau^+\tau^-$\ }
\def\Rs{R \hspace{-0.38em}/\;}
\newc{\beq}{\begin{eqnarray}}
\newc{\eeq}{\end{eqnarray}}
\newc{\dqu}{\delta_{qu}}
\newc{\dqd}{\delta_{qd}}
\newc{\non}{\nonumber}
\newc{\noi}{\noindent}

\def\ib#1,#2,#3{       {\it ibid.\/ }{\bf #1} (19#2) #3}
\def\ap#1,#2,#3{       {\it Ann.~Phys.~(NY)\/ }{\bf #1} (19#2) #3}
\def\ijmp#1,#2,#3{     {\it Int.\ J.~Mod.\ Phys.\/ } {\bf A#1} (19#2) #3}
\def\mpla#1,#2,#3 {     {\it Mod.~Phys.~Lett.\/ } {\bf A#1} (19#2) #3}
\def\npb#1,#2,#3{       {\it Nucl.\ Phys.\/ }{\bf B#1} (19#2) #3}
\def\npps#1,#2,#3{     {\it Nucl.\ Phys.~B (Proc.~Suppl.)\/ }{\bf B#1}
                             (19#2) #3}
\def\plb#1,#2,#3{      {\it Phys.\ Lett.\/ }{\bf B#1} (19#2) #3}
\def\pr#1,#2,#3{       {\it Phys.\ Rev.\/ }{\bf #1} (19#2) #3}
\def\prd#1,#2,#3{      {\it Phys.\ Rev.\/ }{\bf D#1} (19#2) #3}
\def\prep#1,#2,#3{     {\it Phys.\ Rep.\/ }{\bf #1} (19#2) #3}
\def\prl#1,#2,#3{      {\it Phys.\ Rev.\ Lett.\/ }{\bf #1} (19#2) #3}
\def\pro#1,#2,#3{      {\it Prog.~Theor.\ Phys.\/ }{\bf #1} (19#2) #3}
\def\rmp#1,#2,#3{      {\it Rev.~Mod.~Phys.\/ }{\bf #1} (19#2) #3}
\def\sp#1,#2,#3{       {\it Sov.~Phys.~Usp.\/ }{\bf #1} (19#2) #3}
\def\zpc#1,#2,#3{      {\it Z.~Phys.\/ }{\bf C#1} (19#2) #3}
\def\appb#1,#2,#3{     {\it Acta Phys.\ Polon.\/ }{\bf B#1} (19#2) #3}

\begin{document}

\begin{flushright}
  BI-TP 97/29\\ DESY 97-062\\ WUE-ITP-97-024\\[1.7ex] {\tt
    hep-ph/9708272} \\ 
\end{flushright}

\vskip 35pt
\begin{center}
  {\Large \bf $R$-Parity Violating SUSY Signals in\\[2mm]   
Lepton-Pair Production at the  Tevatron }\\[2mm] 

\vspace{5mm} 
{\large J. Kalinowski}$^{1,2}$, 
{\large R. R\"uckl}$^{3,\displaystyle \ast}$, 
{\large H. Spiesberger}$^{4,\displaystyle \ast}$,\\[1.1ex] 
{\large and P.M. Zerwas}$^1$\\[2ex] 
{\em $^1$ Deutsches Elektronen-Synchrotron DESY, D-22607
  Hamburg}\\[1.1ex]  
{\em $^2$ Institute of Theoretical Physics, Warsaw University,
  PL-00681 Warsaw}\\[1.1ex] 
{\em $^3$ Institut f\"ur Theoretische Physik, Universit\"at
  W\"urzburg, D-97074 W\"urzburg}\\[1.1ex] 
{\em $^4$ Fakult\"at f\"ur Physik, Universit\"at Bielefeld, D-33501
  Bielefeld}\\[2ex]


\vspace{1cm} {\bf ABSTRACT}
\end{center}

\begin{quotation}
  In supersymmetric theories with $R$-parity breaking, sleptons can be
  produced in quark-antiquark annihilation at the Tevatron through
  interactions in which two quark fields are coupled to a slepton
  field. If at the same time trilinear slepton-lepton-lepton couplings
  are present, the sleptons can be searched for as resonances in
  $p\bar{p}\rightarrow \tilde{\nu} \rightarrow l^+l^-$ and
  $\tilde{\tau}\rightarrow l\nu $ final states.  Existing Tevatron
  data can be exploited to derive bounds on the Yukawa couplings of
  sleptons to quark and lepton pairs. Similar bounds can also be
  obtained from $e^+e^-$ annihilation to hadrons at LEP2.
\end{quotation}

\vspace*{\fill} \footnoterule {\footnotesize
  \noindent ${}^{\displaystyle \ast}$ Supported by Bundesministerium
  f\"ur Bildung, Wissenschaft, Forschung und Technologie, Bonn,
  Germany, Contracts 05 7BI92P (9) and 05 7WZ91P (0).}

\newpage \renewcommand{\thefootnote}{\arabic{footnote}}


\noindent {\bf 1.} 
Supersymmetric theories may include interactions with
$R$-parity breaking \cite{Rp,Dre}. In one of the possible scenarios, the
superpotential involves terms in which three lepton superfields and
lepton-quark-quark superfields are coupled:
\begin{equation}
  W_{\Rs}=\lambda_{ijk}L_iL_j E^c_k + \lambda'_{ijk}L_iQ_j D^c_k
   \label{superp}
\end{equation} 
The indices $ijk$ denote the generations; the couplings $\lambda_{ijk}$ are
non-vanishing only for $i < j$. 
The left-handed doublets of the leptons are denoted by $L$, the
right-handed singlets by $E$, correspondingly the quark doublets by
$Q$ and the quark singlets by $D$. Both terms violate the lepton number,
yet conserve the baryon number. In four-component Dirac notation the 
Yukawa interactions have the following form
\begin{eqnarray}&& {\cal 
    L}_{\Rs}=\lambda_{ijk}\left[ \ti{\nu}^j_L\bar{e}^k_Re^i_L
  +\overline{\ti{e}}^k_R (\bar{e}^i_L)^c\nu^j_L
  +\ti{e}^i_L\bar{e}^k_R\nu^j_L -\ti{\nu}^i_L\bar{e}^k_Re^j_L
  -\overline{\ti{e}}^k_R (\bar{e}^j_L)^c\nu^i_L
  -\ti{e}^j_L\bar{e}^k_R\nu^i_L \right] + h.c.    \\ 
 &&\mbox{~~~}+\lambda'_{ijk}\left[\left(
  \ti{u}^j_L\bar{d}^k_Re^i_L
  +\overline{\ti{d}}^k_R(\bar{e}^i_L)^cu^j_L
  +\ti{e}^i_L\bar{d}^k_Ru^j_L \right) 
  -\left(\ti{\nu}^i_L\bar{d}^k_Rd^j_L+\ti{d}^j_L\bar{d}^k_R\nu^i_L
  +\overline{\ti{d}}^k_R(\bar{\nu}^i_L)^cd^j_L\right) \right] +
  h.c.  \nonumber 
\end{eqnarray} 
In the second line, the up (s)quark fields in the first parentheses
and/or the down (s)quark fields in the second may be Cabibbo rotated
in the mass-eigenstate basis.  However, as we will discuss mainly
sneutrino induced processes, we will assume the basis in which only
the up sector is mixed ($i.e.$, $NDD^c$ is ``diagonal"). We will add
comments on the other choice where relevant for the discussion.

This scenario can be explored in various processes. The coupling
$\lambda LLE^c$ could give rise to sneutrino production in $e^+e^-$
annihilation at LEP2 [3-7], the coupling $\lambda'
LQD^c$ to squark production in positron-quark collisions at HERA
\cite{pre}. Moreover, this interaction could lead to
the formation of sneutrinos and charged sleptons in quark-antiquark
annihilation at the Tevatron \cite{DH,tev}. Even though
sneutrinos and charged sleptons are expected to have small widths, it
is difficult to extract the signal in the hadronic environment from
jet decays of these particles since the production rate is small and
the background of QCD jets is large.  However, if both interaction
terms are present in the superpotential (\ref{superp}), the sneutrinos
and charged sleptons may decay also to pairs of leptons:
\begin{eqnarray}
&&  p\bar{p} \rightarrow \tilde{\nu} \rightarrow \ell^+\ell^- \label{pair}\\ 
&&  p\bar{p} \rightarrow \tilde{\ell} \rightarrow \ell \nu \label{single}
\end{eqnarray}
If the $R$-parity violating couplings $\lambda,\, \lambda'$ are taken
consistent with current upper limits from low-energy experiments, 
such effects could be observed for sufficiently low masses of
the sleptons. In any case, the existing data for lepton pair
production can be used to derive new interesting bounds on the 
two types of Yukawa couplings.

The phenomenological analysis of the processes (\ref{pair}) and
(\ref{single}) is the main subject of this letter. We will also
include a discussion of the mirror process $e^+e^-\rightarrow
q\bar{q}$, which can be studied in hadron production at
LEP2\footnote{Note that the existence or non-existence of squarks in
  the HERA range \cite{hera} is not linked directly to the production
  of sleptons in the Tevatron and LEP2 range; sleptons are generally
  expected to be lighter than squarks.}.

\noindent {\bf 2.} 
The coexistence of the lepton-number violating couplings $\lambda$ and
$\lambda'$ is not forbidden by the non-observation of proton decay.
However, since they may induce FCNC processes (meson mixing and decays
that are suppressed or forbidden in the SM), strong bounds on these
couplings have been established from low-energy and LEP1 experiments
\cite{limits}.

In SUSY GUT scenarios it is generally expected that the third
generation sleptons and squarks are lighter than sfermions of the
first two generations. Moreover, due to the large mass of the top
quark, the flavor violation might be expected maximal in the third
generation.  Low-energy limits for the third generation sfermions are
not very restrictive. Therefore we will concentrate on effects due to
third generation sleptons. In Ref.~\cite{EFP} the possibility of
$\tilde{\nu}_{\tau} \rightarrow b\bar{b}$ decays, induced by the
$\lambda'_{333}$ coupling, has been discussed in the context of the
process $e^+e^-\rightarrow b\bar{b}$.  However, for analyses at the
Tevatron, the case $\lambda'_{311}\ne 0$ is more interesting since it
allows for sneutrino resonance formation in valence quark collisions.

\setlength{\doublerulesep}{1mm}
\begin{table}[htbp]
\begin{center}
\begin{tabular}{|l|c|c|c|c|}\hline \rule{0mm}{6mm}
Process& Coupling & Limit & For & Ref. 
\\[2mm] \hline\hline \rule{0mm}{6mm}
$K\ra\pi\nu\nu $& $\lambda'_{ijk}$& $\displaystyle 
0.012\, r_{\tilde{d}_R^k}
 $& $j=1,2$ & \cite{AG} 
\\ \rule{0mm}{6mm}
$K\bar{K}$ mixing & $\lambda'_{ijk}$ & $\displaystyle 
0.08\, [r_{\tilde{\nu}_i}^{-2} +r_{\tilde{d}_R^k}
^{-2}]^{-1/4}$ & $j=1,2$ & \cite{AG} 
\\[4mm] \hline \rule{0mm}{6mm}
$\tau\ra\pi\nu_{\tau}$ & $\lambda'_{31k}$ & $\displaystyle 0.16 \,
r_{\tilde{d}_R^k}$& &\cite{BC} \\ \rule{0mm}{6mm}
$D\bar{D}$ mixing & $\lambda'_{ijk}$ & $\displaystyle 0.16\,[
r_{\tilde{l}_i}^{-2} + r_{\tilde{d}_R^k}
^{-2}]^{-1/4}$ & $j=1,2$ &\cite{AG} 
 \\[3mm] \hline 
\end{tabular} \end{center}
\caption{\it Limits relevant for the $\lambda'_{311}$ coupling (see text);
$r_i=m_i/100$ GeV.} 
\label{lim1}
\end{table}

Individual bounds\footnote{The couplings $\lambda'_{1jk}$ derived from
  loop induced neutrino masses are strongly constrained only for large
  $\tan\beta$ \cite{josh}.} relevant for
$\lambda'_{311}$ are summarized in Table~\ref{lim1}. The limits from
$K$ decay and $K\bar{K}$ mixing have been derived for the basis in
which $NDD^c$ is diagonal; if $EUD^c$ is diagonal, limits can be
derived from $D\bar{D}$ mixing.

The bound on the coupling $\lambda_{131}$, which determines the
$\tilde{\nu}_{\tau}\ra e^+e^-$ and $\tilde{\tau}\ra e\nu_e$ decay
rates, has recently been updated to the value $\lambda_{131}\leq 0.08$
for a mass scale of 200 GeV \cite{KRRZ2}. For the coupling
$\lambda_{232}$, which determines the $\tilde{\nu}_{\tau}\ra
\mu^+\mu^-$ and $\tilde{\tau}\ra \mu\nu_\mu$ decay rates, an upper
limit of 0.08 has been established in Ref.~\cite{BGH}.  In summary,
present low-energy data allow the product of couplings relevant for
the processes (\ref{pair}) and (\ref{single}) to be of the order
$\lambda_{i3i}\lambda'_{311}\leq (0.05)^2$ for $i=1$ and 2.  In the
following analysis we will consider a specific scenario in which we
take $\lambda_{131}\lambda'_{311}=(0.05)^2$ to illustrate the impact
of possible sneutrino and charged slepton resonance formation on 
$e^+e^-$ or $e\nu_e$ production at the Tevatron. The same results
hold for $\mu^+\mu^-$ or $\mu\nu_{\mu}$ production if
$\lambda_{232}\lambda'_{311}=(0.05)^2$ is assumed.

\noindent {\bf 3.} 
In supersymmetric theories with $R$-parity breaking of the type
(\ref{superp}), lepton pairs are produced in $p\bar{p}$ collisions, in
addition to the standard Drell-Yan processes, primarily through
$s$-channel sneutrino and charged slepton exchanges if the masses are
small enough, Fig.~1:

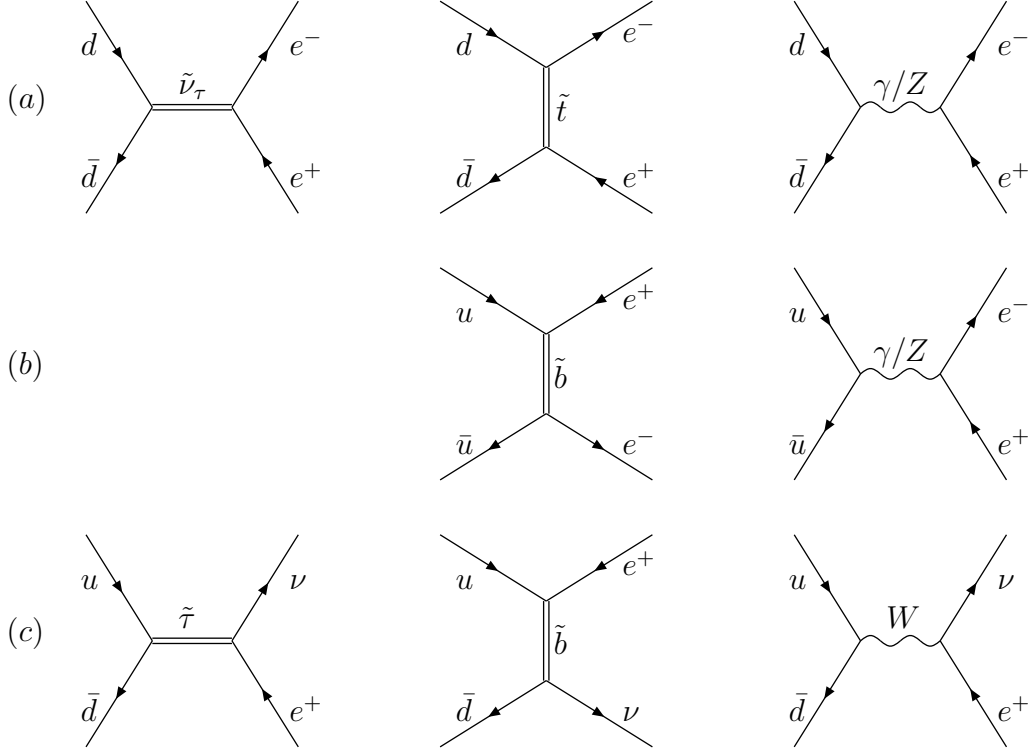
\begin{figure}[htbp]
%
\begin{picture}(100,100)(-20,0)
\ArrowLine(10,90)(35,50)
\ArrowLine(35,50)(10,10)
\Line(35,49)(65,49)
\Line(35,51)(65,51)
\ArrowLine(90,10)(65,50)
\ArrowLine(65,50)(90,90)
\put(8,20){$\bar{d}$}
\put(8,70){$d$}
\put(45,55){$\tilde{\nu}_{\tau}$}
\put(87,20){$e^+$}
\put(87,70){$e^-$}
\put(-20,50){$(a)$}
\end{picture}
%
\begin{picture}(100,100)(-50,0)
\ArrowLine(10,90)(50,65)
\ArrowLine(50,65)(90,90)
\Line(49,65)(49,35)
\Line(51,65)(51,35)
\ArrowLine(90,10)(50,35)
\ArrowLine(50,35)(10,10)
\put(16,20){$\bar{d}$}
\put(16,70){$d$}
\put(54,45){$\tilde{t}$}
\put(79,20){$e^+$}
\put(79,75){$e^-$}
\end{picture}
%
\begin{picture}(100,100)(-80,0)
\ArrowLine(10,90)(35,50)
\ArrowLine(35,50)(10,10)
\Photon(35,50)(65,50){2}{2}
\ArrowLine(90,10)(65,50)
\ArrowLine(65,50)(90,90)
\put(8,20){$\bar{d}$}
\put(8,70){$d$}
\put(40,55){$\gamma/Z$}
\put(87,20){$e^+$}
\put(87,70){$e^-$}
\end{picture}
\\
\begin{picture}(100,100)(-20,0)
\put(-20,50){$(b)$}
\end{picture}
%
\begin{picture}(100,100)(-50,0)
\ArrowLine(10,90)(50,65)
\ArrowLine(90,90)(50,65)
\Line(49,65)(49,35)
\Line(51,65)(51,35)
\ArrowLine(50,35)(90,10)
\ArrowLine(50,35)(10,10)
\put(16,20){$\bar{u}$}
\put(16,70){$u$}
\put(53,45){$\tilde{b}$}
\put(79,20){$e^-$}
\put(79,75){$e^+$}
\end{picture} 
%
\begin{picture}(100,100)(-80,0)
\ArrowLine(10,90)(35,50)
\ArrowLine(35,50)(10,10)
\Photon(35,50)(65,50){2}{2}
\ArrowLine(90,10)(65,50)
\ArrowLine(65,50)(90,90)
\put(8,20){$\bar{u}$}
\put(8,70){$u$}
\put(40,55){$\gamma/Z$}
\put(87,20){$e^+$}
\put(87,70){$e^-$}
\end{picture}
\\
%
\begin{picture}(100,100)(-20,0)
\ArrowLine(10,90)(35,50)
\ArrowLine(35,50)(10,10)
\Line(35,49)(65,49)
\Line(35,51)(65,51)
\ArrowLine(90,10)(65,50)
\ArrowLine(65,50)(90,90)
\put(8,20){$\bar{d}$}
\put(8,70){$u$}
\put(45,55){$\tilde{\tau}$}
\put(87,20){$e^+$}
\put(87,70){$\nu$}
\put(-20,50){$(c)$}
\end{picture}
%
%
\begin{picture}(100,100)(-50,0)
\ArrowLine(10,90)(50,65)
\ArrowLine(90,90)(50,65)
\Line(49,65)(49,35)
\Line(51,65)(51,35)
\ArrowLine(50,35)(90,10)
\ArrowLine(50,35)(10,10)
\put(16,20){$\bar{d}$}
\put(16,70){$u$}
\put(53,45){$\tilde{b}$}
\put(79,20){$\nu$}
\put(79,75){$e^+$}
\end{picture}
%
%
\begin{picture}(100,100)(-80,0)
\ArrowLine(10,90)(35,50)
\ArrowLine(35,50)(10,10)
\Photon(35,50)(65,50){2}{2}
\ArrowLine(90,10)(65,50)
\ArrowLine(65,50)(90,90)
\put(8,20){$\bar{d}$}
\put(8,70){$u$}
\put(45,55){$W$}
\put(87,20){$e^+$}
\put(87,70){$\nu$}
\end{picture}
\caption{\label{drellee} \it Mechanisms in the scenario  
  $\lambda_{131}\lambda'_{311}\ne0$ (first column) and
  $\lambda'_{131}\ne 0$ (second column) for (a) $d\bar{d} \rightarrow
  e^+ e^-$ scattering including $s$-channel exchange of
  $\tilde{\nu}_{\tau}$ and $t$-channel exchange of $\tilde{t}_L$; (b)
  $u\bar{u}\rightarrow e^+e^-$ scattering including $u$-channel
  exchange of $\tilde{b}_R$; and (c) $u\bar{d} \rightarrow e^+\nu$
  scattering including $s$-channel exchange of $\tilde{\tau}$ and
  $u$-channel exchange of $\tilde{b}_R$. The Drell-Yan processes are
  shown in the third column.}
\end{figure}

\noindent The $t$- and/or $u$-channel exchanges of squarks also
contribute to the cross sections; they interfere with the Standard
Model $\gamma,Z$ exchanges. However the Yukawa coupling
$\lambda'_{131}$ is expected to be so small ($\lsim 0.05$) that these
mechanisms will not play an important role in practice\footnote{The
  complete expressions for the cross sections, including the
  contributions from squark exchange, are given in the appendix.}.
Since sleptons are produced in collisions of quarks and antiquarks
with the same helicities while the helicities are opposite in the
Drell-Yan processes, the amplitudes for the slepton signals and the
Drell-Yan backgrounds do not interfere and the total cross sections
are the incoherent sum of the individual cross sections.  For
electron-pair production they read at the parton level
\begin{eqnarray}
\frac{\mbox{d}\hat{\sigma}}{\mbox{d}\cos\theta} [q\bar{q} \ra
 e^+e^-] = \frac{\mbox{d}\hat{\sigma}[{\tilde{\nu}_{\tau}}]}
{\mbox{d}\cos\theta} +
\frac{\mbox{d}\hat{\sigma}[\gamma,Z]}{\mbox{d}\cos\theta}
\label{sigpair}
\end{eqnarray}
The sneutrino and antisneutrino resonances contribute only to
$d\bar{d}$ annihilation
\begin{eqnarray}
\frac{\mbox{d}\hat{\sigma}[{\tilde{\nu}_{\tau}}]}{\mbox{d}\cos\theta}=
\frac{1}{3}\, 
\frac{\pi\alpha^2\hat{s}}{4}\, \frac{(\lambda_{131}\lambda'_{311}/e^2)^2}
{(\hat{s}-m^2_{\tilde{\nu}_{\tau}})^2+\Gamma^2_{\tilde{\nu}_{\tau}}
m^2_{\tilde{\nu}_{\tau}} } \label{snu}
\end{eqnarray}
 The second term in
(\ref{sigpair}) is due to the standard Drell-Yan  $\gamma, Z$ 
exchange diagrams. The factor $\frac{1}{3}$ in front of (\ref{snu})
[and (\ref{tau})] accounts for the matching of the initial
quark/antiquark colors when the standard luminosity distributions are used
later in Eq.~(\ref{lum}).  $\sqrt{\hat{s}}$ is the center-of-mass
energy of the $q\bar{q}$ system.

For electron-neutrino final states, the cross section at the parton
level can be written in a similar form
\begin{eqnarray}
  \frac{\mbox{d}\hat{\sigma}}{\mbox{d}\cos\theta} [u\bar{d} \ra
  e^+\nu_e] =
  \frac{\mbox{d}\hat{\sigma}[{\tilde{\tau}}]}{\mbox{d}\cos\theta} +
  \frac{\mbox{d}\hat{\sigma}[W]}{\mbox{d}\cos\theta}
\end{eqnarray}
where the $\tilde{\tau}$ contribution is given by 
\begin{eqnarray} 
  \frac{\mbox{d}\hat{\sigma}[{\tilde{\tau}}]}{\mbox{d}\cos\theta}=
  \frac{1}{3}\, \frac{\pi\alpha^2\hat{s}}{8}\,
  \frac{(\lambda_{131}\lambda'_{311}/e^2)^2}
  {(\hat{s}-m^2_{\tilde{\tau}})^2+\Gamma^2_{\tilde{\tau}}
    m^2_{\tilde{\tau}} } \label{tau}
\end{eqnarray}
while the second term is due to the Drell-Yan $W$-exchange diagram. 
The angular distribution of the final state leptons is isotropic for
scalar $\tilde{\nu}_{\tau}$ or $\tilde{\tau}$ production in the
slepton center-of-mass frame.  

For $p\bar{p}\ra e^+e^-$ and $e\nu_e$
the differential cross sections are obtained by combining the parton
cross sections with the luminosity spectra for quark-antiquark
annihilation
\begin{eqnarray}
\frac{\mbox{d}^2\sigma}{\mbox{d}M_{\ell\ell}\mbox{d}y}
[p\bar{p}\ra \ell_1\ell_2]=\sum_{ij} \frac{1}{1+\delta_{ij}}
\, \left(f_{i/p}(x_1)  f_{j/\bar{p}}(x_2) +(i 
\leftrightarrow j)\right)\,  \hat{\sigma} [ij\ra \ell_1\ell_2] 
\label{lum}
\end{eqnarray}
where $\ell_1\ell_2=e^+e^-$ or $e^\pm\nu$, $x_1=\sqrt{\tau}e^y$,
$x_2=\sqrt{\tau}e^{-y}$.  $M_{\ell\ell}=(\tau s)^{1/2} =
(\hat{s})^{1/2}$ is the mass and $y$ the rapidity of the lepton pair.
The probability to find a parton $i$ with momentum fraction $x_i$ in
the (anti)proton is denoted by $f_{i/p(\bar{p})}(x_i)$.  When the
cross sections are compared with the data from the CDF Collaboration
\cite{CDF}, the CTEQ3L parametrization \cite{cteq} is used together
with a multiplicative $K$ factor for the higher order QCD corrections
to Drell-Yan pair production; this follows the procedure of
Ref.~\cite{CDF}.  Since the corresponding $K$ factor for slepton
production has not been determined yet, the couplings
$\lambda\lambda'$ are theoretically uncertain at a level 
of about 10\% which is tolerable at the present stage of the analysis.

\begin{figure}[htb] 
 \unitlength 1mm
\begin{picture}(100,100)
  \put(-20,-110){ \epsfxsize=20cm \epsfysize=25cm
\epsfbox{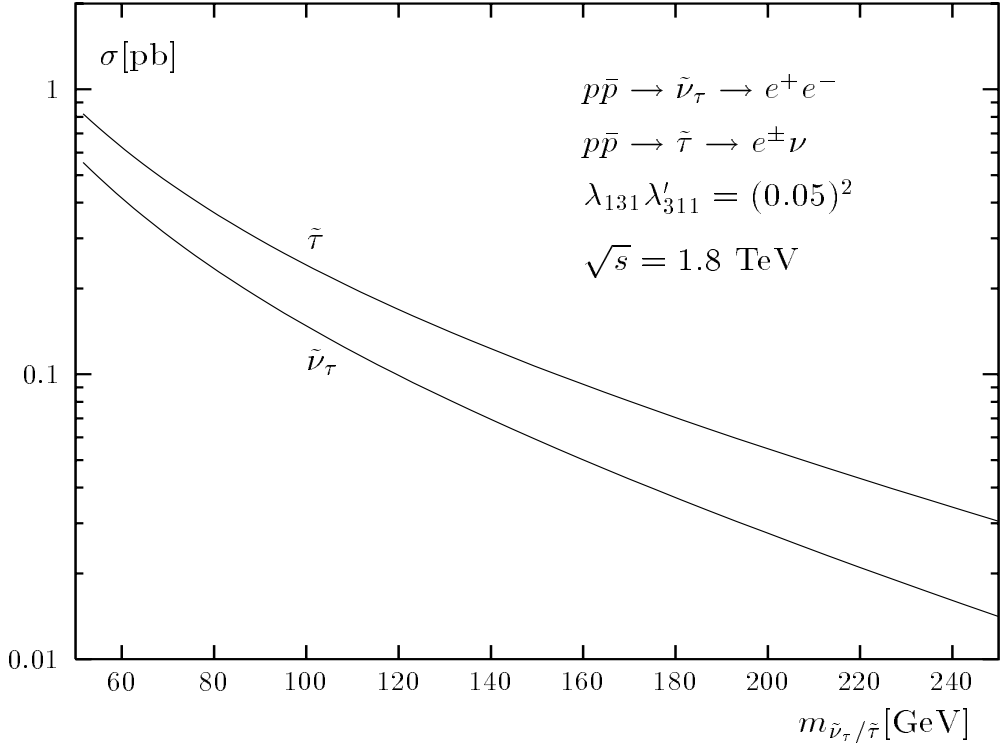}}
\end{picture}
\caption{\it 
  The cross sections for sneutrino ($\tilde{\nu}_{\tau}$) and antisneutrino 
($\;\overline{\tilde{\nu}}_{\tau}$) and stau
  ($\tilde{\tau}$) production at the Tevatron, including branching
  ratios to lepton-pair decays, as a function of the slepton mass. 
  Parameters: $\lambda_{131}\lambda'_{311} = (0.05)^2$,
  $\protect\Gamma_{\tilde{\nu}_{\tau}}=\protect\Gamma_{\tilde{\tau}} =
  1$ GeV.  }
\label{sigt}
\end{figure}

The total cross sections for sneutrino and charged slepton production
in the $e^+e^-$ and $e\nu_e$ channels at the Tevatron are shown as a
function of the slepton mass in Fig.~\ref{sigt}.  The scalar particles are
assumed to belong to the third generation, $\tilde{\tau}$ and
$\tilde{\nu}_{\tau}$, and the two Yukawa couplings are taken to be 
$\lambda_{131}\lambda'_{311}=(0.05)^2$ for illustration, compatible
with the bounds derived at low energies. The widths of
$\tilde{\nu}_{\tau}$ and of $\tilde{\tau}$ have been set to the  typical
value of $\Gamma=1$ GeV, corresponding to branching ratios of ${\cal O} (
0.01)$ for leptonic decays.

\begin{figure}[htb] 
  \unitlength 1mm
\begin{picture}(100,100)
  \put(-20,-110){ \epsfxsize=20cm \epsfysize=25cm
\epsfbox{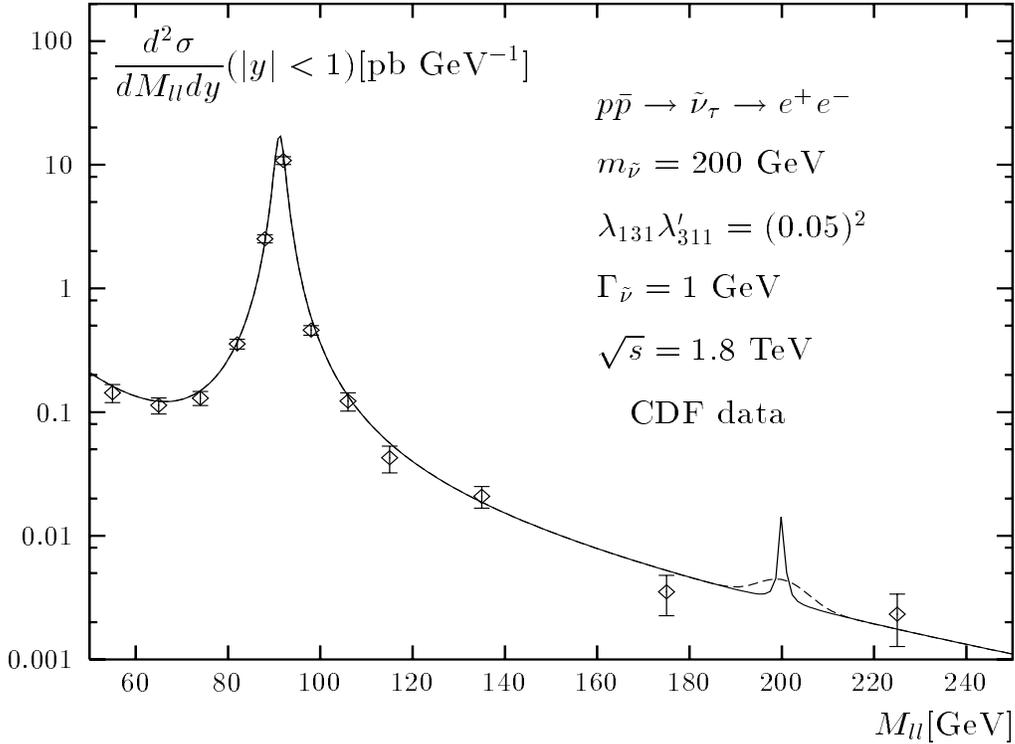}}
\end{picture}
\caption{\it 
  The $\protect e^+e^-$ invariant mass distribution
  $\protect\mbox{\rm d}\sigma/\mbox{\rm d}M_{\ell\ell}\mbox{\rm d}y$
  ($|y|<1$) in $p\bar{p}$ collisions at $\protect\sqrt{s}=1.8$ TeV
  observed by CDF \protect\cite{CDF}, is compared with the theoretical
  prediction of the Drell-Yan mechanism supplemented by the
  $s$-channel exchange of a sneutrino in the channel $\protect
  d\bar{d}\rightarrow e^+e^-$ with $\protect
  m_{\tilde{\nu}_{\tau}}=200 $ GeV, $\protect
  \lambda_{131}\lambda'_{311} = (0.05)^2$ and
  $\protect\Gamma_{\tilde{\nu}_{\tau}}=1$ GeV. Solid line: ideal
  detector, dashed line: sneutrino resonance smeared with a Gaussian
  of width 5 GeV. Parton distributions are taken from CTEQ3L.}
\label{massdist}
\end{figure}

The size of the peak in the distribution of the di-electron invariant
mass $M_{ee}$ is depicted in Fig.~\ref{massdist}. Following again 
CDF practice, the differential cross section has been integrated over
the range $|y|<1$ and subsequently divided by 2 to account for two
units of the rapidity.  While the full curve represents the
distribution for an ideal detector, the dashed curve demonstrates the
smearing of the peak by experimental resolution, characterized by a
Gaussian with a width of 5 GeV.  The transverse momentum distributions
of the electrons in the decay processes $\tilde{\nu}_{\tau}\rightarrow
e^+e^-$ and in $\tilde{\tau}\rightarrow e\nu$ develop Jacobian peaks,
which however are smeared out by the non-zero widths of the sfermions
and QCD radiative corrections in the initial state.

In the same figure the dilepton spectrum for $p\bar{p}\rightarrow
e^+e^-$, measured by the CDF Collaboration at the Tevatron
\cite{CDF}, is compared with the prediction of the Standard Model
modified by the presence of a hypothetical sneutrino resonance at 200
GeV. Assuming the sneutrino contribution to be smaller than the
experimental error of the data points, the product of the Yukawa
couplings can be estimated to be
\begin{equation} 
(\lambda_{131}\lambda'_{311})^{1/2} \lsim 0.08\; \gamma^{1/4} \label{cdflim}
\end{equation}
for sneutrino masses in the range 120 -- 250 GeV.  The decay
width of the sneutrino enters the estimate only weakly through the
fourth root of the parameter
$\gamma$ which denotes the sneutrino width in units of GeV.  

\noindent {\bf 4.} 
The same product $\lambda_{131}\lambda'_{311}$ can also be studied in
$e^+e^-$ annihilation to hadrons at LEP2.  Neglecting the small
contribution from the $t/u$-channel squark exchange the cross section
can be written as the incoherent sum of the sneutrino and the Standard
Model $\gamma,Z$ contributions:
\begin{eqnarray}
\frac{\mbox{d}\hat{\sigma}}{\mbox{d}\cos\theta} [e^+e^- \ra
 q\bar{q}] = \frac{\mbox{d}\hat{\sigma}[{\tilde{\nu_{\tau}}}]}
{\mbox{d}\cos\theta} +
\frac{\mbox{d}\hat{\sigma}[\gamma,Z]}{\mbox{d}\cos\theta}
\end{eqnarray}
The sneutrino and antisneutrino resonances contribute only to the
$d\bar{d}$ final state:
\begin{eqnarray}
\frac{\mbox{d}\hat{\sigma}[{\tilde{\nu}}]}{\mbox{d}\cos\theta}= 
\frac{3\pi\alpha^2s}{4}\, \frac{(\lambda_{131}\lambda'_{311}/e^2)^2}
{(s-m^2_{\tilde{\nu}_{\tau}})^2+\Gamma^2_{\tilde{\nu}_{\tau}}
m^2_{\tilde{\nu}_{\tau}} } 
\end{eqnarray}
\begin{figure}[htb] 
 \unitlength 1mm
\begin{picture}(90,90)
  \put(-20,-155){ \epsfxsize=20cm \epsfysize=30cm
\epsfbox{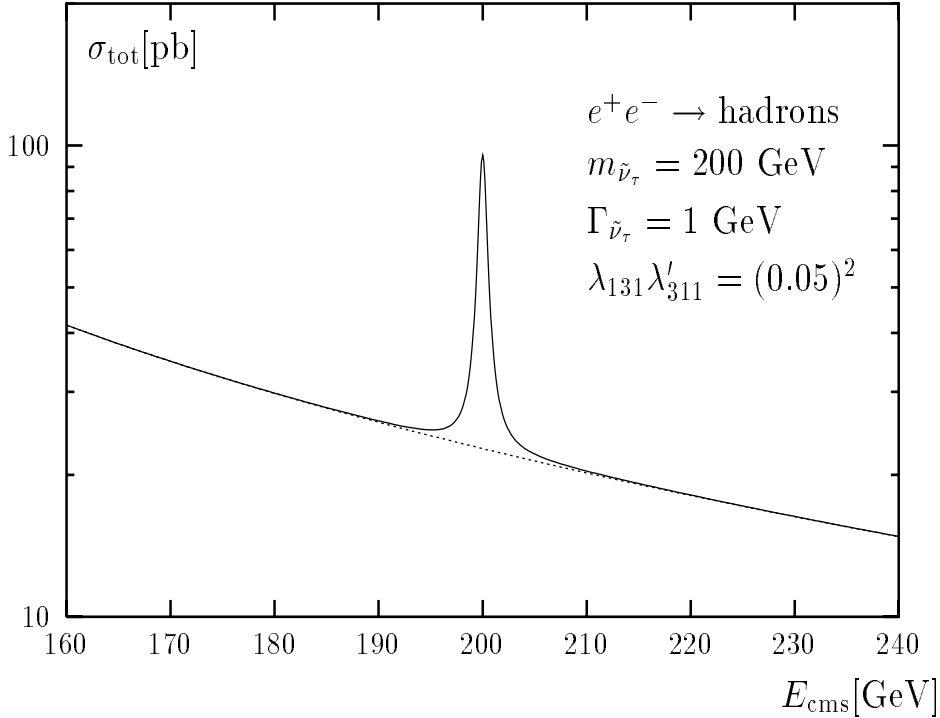}}
\end{picture}
\caption{\it 
  Cross section for the process $ e^+e^-\rightarrow hadrons$ as a
  function of the total energy. The prediction of the Standard Model
  $\gamma,Z$ exchange is given by the dotted line. The contribution of
  a potential $\tilde{\nu}_{\tau}$ (and
  $\overline{\tilde{\nu}}_{\tau}$) resonance is superimposed in the
  full curve.  Parameters: $\lambda_{131}\lambda'_{311} = (0.05)^2$,
  $\protect\Gamma_{\tilde{\nu}_{\tau}}= 1$ GeV.}
\label{sigeedd}
\end{figure}

\noindent The total hadronic cross section for $e^+e^-$ annihilation
is presented in Fig.~\ref{sigeedd}, again for
$\lambda_{131}\lambda'_{311}=(0.05)^2$ and for a sneutrino mass of
$m_{\tilde{\nu}_{\tau}}=200 $ GeV. The angular distribution of the $d$
and $\bar{d}$ jets is nearly isotropic on the sneutrino resonance.  As
a result, the strong forward-backward asymmetry in the Standard Model
continuum, $A_{FB}\sim 0.65$ at $\sqrt{s}=200$ GeV, is reduced to
$\sim 0.03$ on top of the sneutrino resonance.  If the total cross
section $\sigma[{e^+e^-\rightarrow hadrons}]$ can be measured to an
accuracy of about 1\% at $\sqrt{s}=184$ $(192)$ GeV, the Yukawa
couplings for a 200 GeV sneutrino can be bounded to
\begin{equation} 
(\lambda_{131}\lambda'_{311})^{1/2}\lsim 0.072\; (0.045) \label{leplim}
\end{equation}
This bound compares well with the bound Eq.~(\ref{cdflim}) which has
been derived from the Tevatron data. Since the limit (\ref{leplim}) is
estimated for energies much below the resonance,
$|\sqrt{s}-m_{\tilde{\nu}_{\tau}}| \ll \Gamma_{\tilde{\nu}_{\tau}}$,
it is independent of the sneutrino width.  If the light sneutrino
exists, the Yukawa coupling $\lambda_{131}$ can be measured separately
in Bhabha scattering $e^+e^-\rightarrow \tilde{\nu}_{\tau}\rightarrow
e^+e^-$ at LEP2, cf.\ Ref.~\cite{KRRZ2}, which in turn would
allow to derive limits on $\lambda'_{311}$.  

\noindent {\bf 5.} 
In conclusion, the simultaneous presence of lepton number violating
couplings $\lambda$ and $\lambda'$ can manifest itself in a large
number of different reactions. We have shown that taking the product
of couplings $\lambda_{131}\lambda'_{311}$ according to the present
limits from low-energy experiments, an excess in lepton-pair
production at the Tevatron could be observed for sufficiently low
masses of sleptons. Conversely, if no excess of events in $p\bar{p}
\rightarrow e^+e^-$ or $e\nu$ are observed, stringent bounds on
products of the $\lambda\lambda'$ Yukawa couplings can be extracted.
Similar sensitivities to the size of Yukawa couplings are expected in
$e^+e^-$ annihilation to hadrons. Combining these limits with bounds
on $\lambda$ couplings from purely leptonic processes in $e^+e^-$
collisions, individual constraints on $\lambda$ and $\lambda'$
couplings can be established in an experimentally direct way.
\\

\noindent {\Large \bf Acknowledgments}

\noindent We are grateful to G. Bhattacharyya, J. Conway, S.~Eno, 
A.S.~Joshipura and Y.~Sirois for discussions. Communications
by K.~Maeshima and W.~Sakumoto concerning the CDF data points are
gratefully acknowledged.
\\

\noindent {\Large \bf Appendix}

\noindent \underline{a) $q\bar{q}\rightarrow \ell_1\ell_2$}

The differential cross section for the process $q\bar{q}^{(')}\rightarrow
\ell_1\ell_2$ in the quark-antiquark rest frame can be written most
transparently in terms of helicity amplitudes 
\begin{eqnarray}
&&\frac{\mbox{d}\hat{\sigma}}{\mbox{d}\cos\theta}[q\bar{q}^{(')}
\rightarrow \ell_1\ell_2]
 = \non 
\frac{\pi\alpha^2\hat{s}}{24} \Bigl\{
4\left[|f^t_{LL}|^2+|f^t_{RR}|^2\right] \\[1.5ex]
&& + (1+\cos\theta)^2 
\left[ |f^s_{LR}|^2 + |f^s_{RL}|^2 +
  |f^t_{LR}|^2 + |f^t_{RL}|^2 + 2\mbox{Re}(f^{s\;*}_{LR}\,f^t_{LR}) 
  + 2\mbox{Re}(f^{s\;*}_{RL}\,f^t_{RL}) \right] \non \\[1.5ex]
&&+ (1-\cos\theta)^2\left[|f^s_{LL}|^2+|f^s_{RR}|^2\right] 
 \Bigr\}
\label{sigbb}
\end{eqnarray}
The complete cross section for $q\bar{q}\rightarrow e^+e^-$ is built
up by the $s$-channel exchange of $\gamma,Z$ bosons and by additional
contributions due to sfermion exchanges: the $s$-channel
$\tilde{\nu}_{\tau}$ and the $t$-channel $\tilde{t}_L$ exchanges in
$d\bar{d}\rightarrow e^+e^-$, and the $u$-channel $\tilde{b}_R$
exchange in the $u\bar{u}\rightarrow e^+e^-$ subprocess (see
Fig.~\ref{drellee}a, b).

While the $s$-channel $\gamma,Z$ amplitudes in the Standard Model
involve the coupling of vector currents, the sfermion exchange is
described by scalar densities. By performing appropriate Fierz
transformations, the $s$-channel $\ti{\tau}$ exchange amplitudes can
formally be rewritten as $t$-channel vector amplitudes, and
$t/u$-channel $\ti{q}$ exchange amplitudes as $s$-channel vector
amplitudes \cite{KRRZ1}; for the operators: $(\bar{f}_R
f'_L)(\bar{F}_L F'_R) \ra -\frac{1}{2}(\bar{f}_R\gamma_{\mu}
F'_R)(\bar{F}_L\gamma_{\mu}f'_L)$. The independent $s$-channel
amplitudes $f^s_{h_i h_f}$ for $q\bar{q}\rightarrow e^+e^-$ are
therefore given by 
\begin{eqnarray} 
f^s_{LR} &=& - \frac{Q^q}{\hat{s}} + \frac{g_L^q
  g_L^e}{\hat{s}-m^2_Z+i\Gamma_Zm_Z}
\label{fslr} \\ 
f^s_{RL} &=& -\frac{Q^q}{\hat{s}} 
             + \frac{g_R^q g_R^e}{\hat{s}-m^2_Z+i\Gamma_Zm_Z}
\label{fsrl} \\ 
f^s_{LL} &=& -\frac{Q^q}{\hat{s}} 
             + \frac{g_L^qg_R^e}{\hat{s}-m^2_Z+i\Gamma_Zm_Z}
-\frac{1}{2}\, \frac{(\lambda'_{131}/e)^2}{\hat{u}-m^2_{\tilde{b}_R}} 
\delta_{qu}  \label{fsll} \\ 
f^s_{RR} &=& - \frac{Q^q}{\hat{s}} 
             +\frac{g_R^q g_L^e}{\hat{s}-m^2_Z+i\Gamma_Zm_Z}
+\frac{1}{2}\, \frac{(\lambda'_{131}/e)^2}{\hat{t}-m^2_{\tilde{t}_L}} 
\delta_{qd} 
\label{fsrr}
\end{eqnarray}
where $\hat{t} = - \hat{s}(1-\cos \theta)/2$, $\hat{u} = -
\hat{s}(1+\cos \theta)/2$; the $t$-channel amplitudes $f^t_{h_i
  h_f}$ read 
\begin{eqnarray}
f^t_{LR} &=&0 \label{ftlr} \\
f^t_{RL} &=&0 \label{ftrl} \\
f^t_{LL}&=& \frac{1}{2}\, \frac{\lambda_{131}\lambda'_{311}/e^2}
{\hat{s}-m^2_{\tilde{\nu}_{\tau}}+i\Gamma_{\tilde{\nu}_{\tau}}
m_{\tilde{\nu}_{\tau}} }\delta_{qd}  \label{ftll} \\
f^t_{RR}&=& \frac{1}{2}\, \frac{\lambda_{131}\lambda'_{311}/e^2}
{\hat{s}-m^2_{\tilde{\nu}_{\tau}}+i\Gamma_{\tilde{\nu}_{\tau}}
m_{\tilde{\nu}_{\tau}} }\delta_{qd}  \label{ftrr}
\end{eqnarray}
Note the relative sign between the $\tilde{b}$ and $\tilde{t}$
contributions due to the different ordering of fermion operators 
in the Wick reduction. To
simplify the notation, we have defined the indices $L,R$ to denote the
helicities of the {\it ingoing quark} (first index) and of the {\it
  outgoing positron} (second index).  The helicities of the ingoing
antiquark and the outgoing electron are fixed by the $\gamma_5$
invariance of the vector interactions: they are opposite to the
helicities of the quark and positron in $s$-channel amplitudes and the
same in $t$-channel amplitudes. The left/right $Z$
charges\footnote{Note that in Eqs.~(\ref{fslr}-\ref{ftrr}) the
  outgoing positron with the helicity $L(R)$ couples with the charge
  $g_R (g_L)$.} of the fermions are defined as
\begin{eqnarray}
g^f_L=\left(\frac{\sqrt{2}G_{\mu}m^2_Z}{\pi\alpha}\right)^{1/2}
\left[I_3^f-s^2_W Q^f\right] \mbox{~~~and~~~}  
g^f_R=\left(\frac{\sqrt{2}G_{\mu}m^2_Z}{\pi\alpha}\right)^{1/2}
\left[  {} -s^2_W Q^f\right] \label{gis}
\end{eqnarray}
where $s^2_W=\sin^2\theta_W$.

The analysis of the process $q\bar{q}'\rightarrow e\nu$,  
Fig.~\ref{drellee}c, 
proceeds analogously. The helicity amplitudes for
$u\bar{d}\rightarrow e^+\nu_e$ can be written as follows: 
\begin{eqnarray}
f^s_{LR} &=& \frac{1}{2s^2_W}\, \frac{1}{\hat{s}-m^2_W+i\Gamma_W m_W} \\ 
f^s_{LL} &=&-\frac{1}{2}\,
\frac{(\lambda'_{131}/e)^2}{\hat{u}-m^2_{\tilde{d}_R}} \\ 
f^t_{LL} &=&-\frac{1}{2}\, \frac{\lambda_{131}\lambda'_{311}/e^2}
{\hat{s}-m^2_{\tilde{\tau}}+i\Gamma_{\tilde{\tau}} m_{\tilde{\tau}}} 
\end{eqnarray}
while all other helicity amplitudes vanish.\\

\noindent \underline{ b) $e^+e^-\rightarrow q\bar{q}$}

For completeness we include the explicit expressions for the process
$e^+e^-\rightarrow q\bar{q}$ (the contributing Feynman diagrams are
the time-reversed versions of the diagrams in Fig.~\ref{drellee}a, b).
The cross section can be written as
\begin{eqnarray}
&&\frac{\mbox{d}\hat{\sigma}}{\mbox{d}\cos\theta}[e^+e^-\rightarrow q\bar{q}]
 = \non 
\frac{3\pi\alpha^2s}{8} \Bigl\{
4\left[|f^t_{LL}|^2+|f^t_{RR}|^2\right] \\[1.5ex]
&& + (1+\cos\theta)^2 
\left[ |f^s_{LR}|^2 + |f^s_{RL}|^2 +
  |f^t_{LR}|^2 + |f^t_{RL}|^2 + 2\mbox{Re}(f^{s\;*}_{LR}\,f^t_{LR}) 
  + 2\mbox{Re}(f^{s\;*}_{RL}\,f^t_{RL}) \right] \non \\[1.5ex]
&&+ (1-\cos\theta)^2\left[|f^s_{LL}|^2+|f^s_{RR}|^2\right] 
 \Bigr\}
\label{diff-xsec} 
\end{eqnarray}
The indices $L,R$ in the helicity amplitudes $f$ denote the helicity
of the {\it incoming electron} (first index) and of the {\it outgoing
  antiquark} (second index). Again, after applying  Fierz
transformations, the amplitudes can be cast in the following form:
\begin{eqnarray} 
f^s_{LR} &=& -\frac{Q^q}{s} +
\frac{g_L^e g_L^q}{s-m^2_Z+i\Gamma_Zm_Z}
\label{gslr}   \\ 
f^s_{RL} &=& -\frac{Q^q}{s} + \frac{g_R^e g_R^q}{s-m^2_Z+i\Gamma_Zm_Z}
\label{gsrl} \\ 
f^s_{LL} &=& - \frac{Q^q}{s} + \frac{g_L^eg_R^q}{s-m^2_Z+i\Gamma_Zm_Z}
-\frac{1}{2}\, \frac{(\lambda'_{131}/e)^2}{u-m^2_{\tilde{b}_R}} 
\delta_{qu}  \label{gsll} \\ 
f^s_{RR} &=& -
\frac{Q^q}{s} +\frac{g_R^e g_L^q}{s-m^2_Z+i\Gamma_Zm_Z}
+\frac{1}{2}\, \frac{(\lambda'_{131}/e)^2}{t-m^2_{\tilde{t}_L}} 
\delta_{qd}\label{gsrr} \\
f^t_{LR} &=&0 \label{gtlr} \\
f^t_{RL} &=&0 \label{gtrl} \\
f^t_{LL}&=& \frac{1}{2}\, \frac{\lambda_{131}\lambda'_{311}/e^2}
{s-m^2_{\tilde{\nu}_{\tau}}+i\Gamma_{\tilde{\nu}_{\tau}}
m_{\tilde{\nu}_{\tau}} }\delta_{qd} \label{gtll} \\
f^t_{RR}&=& \frac{1}{2}\, \frac{\lambda_{131}\lambda'_{311}/e^2}
{s-m^2_{\tilde{\nu}_{\tau}}+i\Gamma_{\tilde{\nu}_{\tau}}
m_{\tilde{\nu}_{\tau}} }\delta_{qd} \label{gtrr}
\end{eqnarray}
The gauge couplings $g_{L,R}^{e,q}$ have been defined in Eq.~(\ref{gis}).

\end{document}